\documentclass[reprint %
 reprint,
 amsmath,amssymb,
 aps,
pra,
]{revtex4-2}
\usepackage{float}
\usepackage[T1]{fontenc}
\usepackage{textcomp}
\usepackage{mathptmx}
\usepackage[scaled=.92]{helvet}
\usepackage{courier}
\usepackage{booktabs} 
\usepackage{graphicx}
\usepackage{bm}
\usepackage{newtxtext}
\usepackage{newtxmath}
\usepackage{natbib}
\usepackage{hyperref}
\usepackage{cleveref}
\usepackage{enumitem}   

\newcommand{\spdif}[2]{ \frac{\partial^2 #1}{\partial #2^2}}
\newcommand{\pdif}[2]{ \frac{\partial #1}{\partial #2}}
\newcommand{\na}{ \nabla}  
\newcommand{\eps}{\varepsilon}
\newcommand{\Pe}{P\kern-.06em e}
\usepackage{bm}
\usepackage{letltxmacro}
\LetLtxMacro{\originaleqref}{\eqref}
\renewcommand{\eqref}{Eq.~\originaleqref}
\renewcommand*{\eqref}[1]{Eq.\,\originaleqref{#1}}

\begin{document}

\preprint{APS/123-QED}

\title{A shear-induced limit on bacterial surface adhesion in fluid flow
}

\author{Edwina F. Yeo}
 \email{Contact: edwina.yeo.14@ucl.ac.uk}

\affiliation{Department of Mathematics, University College London, London, WC1H 0AY, UK}

\author{Benjamin J. Walker}
\affiliation{Department of Mathematics, University College London, London, WC1H 0AY, UK}

\author{Philip Pearce}
\affiliation{Department of Mathematics, University College London, London, WC1H 0AY, UK}
\affiliation{Institute for the Physics of Living Systems, University College London, London, WC1H 0AY, UK}

\author{Mohit P. Dalwadi}
\affiliation{Mathematical Institute, University of Oxford, Oxford, OX2 6GG, UK}
\affiliation{Department of Mathematics, University College London, London, WC1H 0AY, UK}


\begin{abstract} 
Controlling bacterial surface adhesion and subsequent biofilm formation in fluid systems is crucial for the safety and efficacy of medical and industrial processes.
Here, we theoretically examine the transport of bacteria close to surfaces, isolating how the key processes of bacterial motility and fluid flow interact and alter surface adhesion.
We exploit the disparity between the fluid velocity and the swimming velocity of common motile bacteria and, using a hybrid asymptotic-computational approach, we systematically derive the coarse-grained bacterial diffusivity close to surfaces as a function of swimming speed, rotational diffusivity, and shape. 
We calculate an analytical upper bound for the bacterial adhesion rate by considering the scenario in which bacteria adhere irreversibly to the surface on first contact. Our theory predicts that maximal adhesion occurs at intermediate flow rates: at low flow rates, increasing flow increases surface adhesion, while at higher flow rates, adhesion is decreased by shear-induced cell reorientation.
           
\end{abstract}
\maketitle

\section{Introduction}\label{sec:headings}

Almost two-thirds of recorded bacteria are motile: able to propel themselves in search of nutrients and adhere to surfaces \cite{wei2011population}. After attachment to a surface,  bacteria can form dense colonies known as biofilms, which are the cause of the fouling of products in food processing, the contamination of drinking water supplies and more than half of all healthcare-associated infections \cite{shirtliff2009role,camara2022economic}. In many of these industrial and natural environments, bacterial transport during the adhesion process is dominated by fluid flow. Despite intensive study, the overall effect of flow on bacterial surface attachment is not known; experimental studies have found that increasing flow rates can either increase or decrease bacterial surface colonisation, depending on the bacterial species, the surface properties and the microfluidic setup \cite{alves2020analysing,moreira2014effects,lecuyer2011shear,palalay2023shear,saur2017impact,park2011effect}. 
 For passive particles, such as metal or plastic nanoparticles, it is well-known that higher mass transport occurs at higher flow rates, which in turn generates greater surface adhesion \cite{kuo1980particle}. By contrast, during bacterial surface adhesion, the biophysical cellular processes that regulate attachment (and, sometimes, subsequent detachment) are altered by flow
\citep{lecuyer2011shear}. In many cases, adhesion rate per cell decreases with higher shear forces \cite{dickinson1995analysis,lecuyer2011shear}. However, species including \textit{Escherichia coli} and \emph{Pseudomonas aeruginosa} have been found to exhibit increased cellular surface adhesion forces in increased shear, because of shear-sensitive catch bonds \cite{thomas2004shear} and the effect of flow-induced reorientation on contact area \citep{palalay2023shear}, respectively. Overall, the specific contributions of cell motility, fluid transport, surface chemistry and bacterial appendages to bacterial surface adhesion have not been identified.

As a step towards addressing this complex problem, we characterise how fluid flow affects bacterial surface attachment in the simplified scenario where bacteria adhere instantly and irreversibly with the surface upon first contact. The irreversible surface adhesion of passive particles can be predicted using an adhesion rate derived from classic Lévêque theory, also referred to as the Smoluchowski-Levich (SL) approximation \citep{busscher2006microbial,levêque1928lois,levich1963physicochemical}. This theory predicts a particle adhesion rate of $J(x) \sim0.538  c_{\infty}(D^{2}\dot{\gamma} /
x)^{1/3}$ at a downstream distance $x$ from the source of the particles, where $D$ is the particle diffusivity, $c_{\infty}$ is the bulk concentration of particles and $\dot{\gamma}$ is the shear rate of the flow, with the prefactor calculated using boundary layer analysis. This classic result predicts that passive particle adhesion scales with $\dot{\gamma}^{1/3}$ as the flow rate increases, capturing the effect of increased mass transport.
However, the adhesion of active bacteria to surfaces cannot in general be described by this classic adhesion rate; bacterial dispersivity does not arise from thermal fluctuations but rather from motility, the interaction of which with fluid flow can produce enhanced non-isotropic diffusion absent from Lévêque's analysis \cite{levêque1928lois}. 
When suspended in an otherwise quiescent fluid, the effective diffusivity of a collection of bacteria with swimming speed $\mathcal V_s$ and rotational diffusion coefficient $D_r$ is isotropic with coefficient $D_{\text{quies}}\approx \mathcal{V}_s^2/2D_r$ in 2D \cite{berg1993random}. The effect of combining motility and
flow, in the absence of adhesion, has been analysed using generalized Taylor dispersion theory (GTD) \citep{frankel1991generalized}. This theory provides an effective diffusion tensor and velocity vector that are both spatially uniform (although potentially anisotropic) by averaging over local changes in orientation and position and taking the long-time limit of bacterial movement. In the high-flow environment we consider, there are large spatial gradients in cell alignment close to the surface, so the classical assumptions of spatial homogeneity required to derive GTD models break down \cite{bearon2003extension,Hill2002}. This means previously calculated GTD diffusion tensors cannot be readily applied in combination with the Lévêque solution to predict bacterial adhesion in flow.

 In this paper, we theoretically address this fundamental problem of bacterial surface adhesion in flow by extending Lévêque's analysis to include the additional complexities of bacterial motility. We first consider a collection of individual bacteria using an agent-based framework, before systematically upscaling this to a continuum description of dilute active suspensions \cite{saintillan2015theory}. 
  Bacterial adhesion results in a diffusive boundary layer where bacterial density is depleted, in which we derive analytical solutions for both the bacterial orientation and density. 
 We identify the maximum theoretical adhesion rate of motile bacteria to surfaces, in the absence of surface chemistry effects or imperfect adhesion events. This adhesion rate increases as the flow rate increases for low shear, reaches a maximal adhesion at intermediate shear, and beyond this the adhesion rate 
is reduced by shear-induced bacterial reorientation.  
These results can be applied to both spherical and elongated bacteria in settings where the flow rate exceeds the swimming speed, such as urinary catheters, the human gut and slow-flowing rivers.
\section{Results}
\subsection{Agent-based model for bacterial transport and adhesion}\label{section:agent}
\begin{figure}[h]
    \centering
    \includegraphics[width=0.7\linewidth]{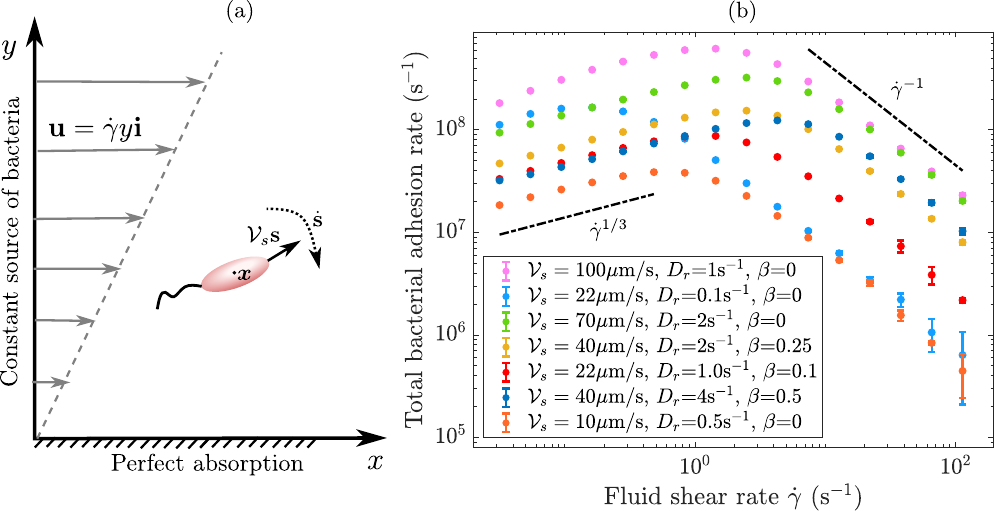}
    \caption{Agent-based simulations of bacterial adhesion in shear flow. (a) Individual bacterium transport in shear flow: each bacterium is advected by the fluid transport $\bm{u}$ and swims with speed $\mathcal{V}_s$ in direction $\bm{s}$, which evolves according to \eqref{SDE-angle}.  (b) Total bacterial adhesion rate as the shear rate $\dot{\gamma}$ varies, with standard deviation across simulations denoted via errorbars (which are very small for lower shear rates). At low shear, the adhesion rate increases as $J\sim \dot{\gamma}^{1/3}$ and, at high shear, the adhesion rate decreases as $J\sim \dot{\gamma}^{-1}$, with analysis of these scalings provided in \S\ref{adhesion-result}. }
    \label{fig:agent}
\end{figure}

We consider the overdamped transport of a dilute active bacterial suspension in a two-dimensional $(x,y)$-plane with an imposed  shear flow $\bm{u}=\dot{\gamma}y\bm{i}$, where $\dot{\gamma}$ is the shear rate and $\bm{i}$ is the unit vector in the $x$-direction. We consider the region $x, y>0$, as shown in Fig.\,\ref{fig:agent}a. We model the bacteria as spheroidal, rigid bodies and consider suspensions dilute enough to neglect physical and hydrodynamic interactions between multiple bacteria, and between bacteria and the surface.   We assume that the bacteria irreversibly adhere to the surface upon first contact. Each bacterium is described by its position vector $\bm{x}$ and unit orientation vector $\bm{s}$, which evolves stochastically, capturing the random changes in direction of the bacteria. The bacteria are advected by the fluid flow and propel themselves in the direction of their orientation vector with fixed speed 
$\mathcal{V}_s$. Hence each bacterium's position is described by a Langevin equation:
\begin{align}
\mathrm{d}\bm{x}=(\bm{u} + \mathcal{V}_s\bm{s})\mathrm{d}t. \label{SDE-vel}
\end{align}
The orientation vector $\bm{s}$ evolves as the bacteria are rotated by the fluid flow, an effect that we refer to as shear alignment, and diffuses randomly at a rate proportional to the rotational diffusion coefficient $D_r$. The fluid flow causes spherical bacteria to rotate at a uniform rate proportional to the rate of rotation tensor $\bm{W}=(\na\bm{u}-\na{\bm{u}}^T)/2$. By contrast, non-spherical bacteria experience non-uniform rotation due to the fluid flow, which results in the classical Jeffery-orbit. This effect is captured by the addition of a term proportional to the rate of strain tensor $\bm{E}=(\na\bm{u}+\na{\bm{u}}^T)/2$ \citep{jeffery1922motion}. Hence, each bacterium's orientation is described by an additional Langevin equation:
\begin{align}
\mathrm{d}\bm{s}=((\beta\bm{E}+\bm{W} ) \bm{s} )\mathrm{d}t+\sqrt{2D_r}\mathrm{d}\bm{\mathcal W} \times \bm{s},\label{SDE-angle}
\end{align}
where the vector $\mathrm{d}\bm{\mathcal W}$ is a Wiener process with mean zero and standard deviation $\sqrt{\mathrm{d}t}$. In \eqref{SDE-angle},  bacterial shape is captured via the Bretherton parameter $\beta\in[0,1)$, which is zero for spherical bacteria and approaches one for infinitely elongated bacteria. 

The adhesion of different bacterial species at different fluid flow rates is shown in Fig.\,\ref{fig:agent}b (numerical framework presented in Appendix \ref{agent-num}). We also vary the motility parameters $\mathcal V_s$ and $D_r$ between species and report adhesion as a function of fluid shear rate $\dot{\gamma}$.
The computed adhesion rates across all motility parameters increase with flow rate at low shear, with the adhesion rate appearing to increase proportionally to $\dot{\gamma}^{1/3}$. This agrees with the scaling predicted by Lévêque for passive particles in flow \cite{levêque1928lois}.
 However, as shear increases further, the adhesion reaches a maximum value before decreasing. The shear rate that generates maximal adhesion differs for each set of bacterial parameters.  These results demonstrate that, even with perfect surface adhesion, there appears to be a maximal rate of bacterial adhesion to surfaces in fluid flow.
 
\subsection{Continuum model for bacterial transport and adhesion}
To understand these observations, we upscale the individual dynamics governed by Eqs.\,(\ref{SDE-vel})--(\ref{SDE-angle}) to describe the dynamics of a collection of bacteria using a continuum partial differential equation (PDE) model, which is amenable to analysis. We use the continuum active suspension model presented in \citep{saintillan2015theory}, and present the details of this upscaling procedure in Supplementary Information \S \ref{S1}. By performing a multiscale (asymptotic) analysis of the continuum PDE model near the surface, we will systematically investigate and explain why adhesion exhibits a maximum as the flow increases, and subsequently calculate the flow rate at which this maximum occurs for given bacterial species.

For ease of analysis, we nondimensionalise our model using a reference velocity scale $\mathcal{U}=\dot{\gamma}\mathcal{L}$ for the external flow, with reference lengthscale $\mathcal{L}$ that is much larger than the size of an individual bacterium, and the timescale of fluid transport $\dot{\gamma}^{-1}$. On the continuum scale, instead of tracking the position of each bacterium we now track three continuum quantities. These are: (i) the density of the collection $\rho(\bm{x},t)$, which captures the number of bacteria per unit area. (ii) The mean orientation vector, known as the polar order parameter $\bm{n}(\bm{x},t)$, which is the continuum equivalent of the orientation vector $\bm{s}$. (iii) The angular distribution of the collection through the $\bm{Q}(\bm{x},t)$ tensor, also known as the nematic order tensor. This tensor can be considered a measure of how locally ordered the bacteria are, noting that $\bm{Q}=\bm{0}$ when the angular distribution of the collection is uniform.
The bacterial density evolves according to a conservation law, in the form of the following PDE:
\begin{eqnarray}
     \frac{D\rho}{Dt}&=&-V_s\na\cdot(\rho\bm{n}),\label{rho_eqn}
    \end{eqnarray}
where the material derivative on the left-hand side captures the advection of the bacteria by the flow and the flux term on the right-hand side captures the swimming of the collection in the direction $\bm{n}.$ The dimensionless parameter $V_s=\mathcal{V}_s/\dot{\gamma}\mathcal{L}$ is the ratio of the bacterial swimming speed to the flow speed. 
    The mean orientation vector $\bm{n}$ and nematic order tensor $\bm{Q}$ capture the bacterial rotation described in \eqref{SDE-angle}, and are described by the following PDEs:
    \begin{subequations}
        \begin{eqnarray}
           \frac{D(\rho\bm{ n})}{Dt}& =&-V_s\left(\na\cdot(\rho\bm{Q})+\frac{1}{2}\na\rho\right) +(\rho\bm{I} \bm{n}-\bm{T}):(\beta \bm{E}+\bm{W})-\frac{\rho\bm{n}}{\Pe_r},\label{n_eqn}
\\
    \frac{D(\rho\bm{Q})}{Dt}&=&-V_s\left(\na\cdot\bm{T}+\frac{\bm{I}}{2}\na\cdot(\rho\bm{n})\right)+
           \beta\rho\left(\bm{E}(\bm{Q}+\bm{I}/2)+(\bm{Q}+\bm{I}/2)\bm{E}\right)+\rho(\bm{W}\bm{Q}-\bm{Q}\bm{W})-2\beta\bm{G}: \bm{E}-\frac{4\rho\bm{Q}}{\Pe_r}.\label{q_eqn}
\end{eqnarray}\label{director-full}
    \end{subequations}
On the left-hand side of both Eqs.\,(\ref{director-full}), the material derivative captures the effect of flow advecting bacteria from upstream.  On the right-hand sides: the flux terms, each premultiplied by the swimming speed $V_s$, capture the effect of bacterial swimming while the source terms featuring the fluid tensors $\bm{W}$ and $\bm{E}$ and the identity tensor $\bm{I}$ capture shear alignment. The last two source terms on the right-hand sides capture rotational diffusion, which is inversely proportional to the rotational P\'eclet number, $\Pe_r=\dot{\gamma}/D_r$, the ratio of fluid rotation to rotational diffusion. There are two additional tensors introduced in Eqs.\,(\ref{director-full}): $\bm{T}$, which is rank three, and $\bm{G}$, which is rank four. These tensors capture higher-order effects that arise when systematically upscaling collective behaviour. We relate these tensors to the mean orientation and nematic order tensor using approximations known as closures. In 2D, these tensors have components defined in index notation as follows \citep{theillard2019computational,fylling2024multi}:
\begin{subequations}
\begin{eqnarray}
\bm{T}_{ijk}&=&\frac{\rho}{4}(\delta_{ij}n_{k}+\delta_{ik}n_{j}+\delta_{jk}n_i),\label{def_ppp}\\
\bm{G}_{ijkl}&=&\frac{\rho}{8}(\delta_{ij}\delta_{lk}+\delta_{ik}\delta_{jl}+\delta_{il}\delta_{jk})+\frac{\rho}{6}(\delta_{ij}Q_{lk}+\delta_{ik}Q_{jl}+\delta_{il}Q_{jk}+\delta_{jk}Q_{il}+\delta_{jl}Q_{ik}+\delta_{jk}Q_{ij}),\label{def_pppp}\end{eqnarray} \label{closures}
\end{subequations}
where $\delta_{ij}$ is the Kronecker delta.
 We demonstrate that these closures, Eqs.\,(\ref{closures}), are very accurate at capturing the underlying agent-based model when: $\Pe_r\lesssim
1$ for highly elongated swimmers $\beta \in(0.5,1)$,  $\Pe_r\lesssim
10$ for modestly elongated swimmers $\beta \in(0,0.5)$ and $\Pe_r\lesssim100$ for circular bacteria (see Supplementary Information \S \ref{validity} for details). Because we do not have hydrodynamic coupling in our model, the fluid flow here is not affected by the bacterial dynamics. We impose a constant bacterial density, $\rho = 1$, as the inlet condition on \eqref{rho_eqn}.  We discuss the inlet conditions for Eqs.\,(\ref{director-full}) in the section below. On the bottom surface at $y=0$ we impose perfect adhesion i.e. $\rho=0$; for no-flux boundaries a careful discussion of boundary conditions is presented in \citep{saintillan2015theory,Maretvadakethope2022}.
 
In this model framework, the behaviour of the collection of bacteria is characterised by two sets of dimensional parameters. The first set are the fluid parameters: the shear rate of the flow $\dot{\gamma}$ and the lengthscale of the flow $\mathcal{L}$, which together set the velocity scale $\mathcal{U}=\dot{\gamma}\mathcal{L}$. Depending on the application of interest, natural choices for the flow lengthscale $\mathcal{L}$ include the radius of the transporting pipe and the size of the viscous fluid boundary layer. We focus on applications in which the fluid speed is significantly faster than the speed of an individual bacterium, and hence the relative swimming speed is small, $V_s\ll1$. Mathematically, this is a singular limit, and we will exploit this in our following analysis. Finally, we note that our analysis and subsequent results are not limited to situations with exact shear flow. Our results will also apply to general flows which can be well approximated by shear flow within the diffusive boundary layer that we identify and analyse below.

\subsubsection*{Transport far from the surface}\label{outer}
We start by briefly summarising the behaviour of the suspension in the flow far from the surface, giving full details in \S \ref{outer-section} of the Supplementary Information. In the next section we discuss how this behaviour changes close to the surface. 
Far from the surface, the bacterial density described by \eqref{rho_eqn} is unaffected by adhesion. Since the relative swimming speed is small ($V_s\ll1$), the bacterial distribution is determined here by fluid transport alone: 
\begin{align}
    \frac{D\rho}{Dt}&=0.\label{rho0_eqn}
\end{align}
\eqref{rho0_eqn} is a statement that bacteria move along fluid streamlines. Hence, bacterial density is constant throughout the flow far from the surface, and is set by the upstream value $\rho=1$. 
In the absence of swimming, the mean bacterial orientation $\bm{n}$ and the nematic order tensor $\bm{Q}$ on each streamline are determined by fluid transport, shear alignment and rotational diffusion alone. 
In the small swimming limit $V_s\ll1$, Eqs.\,(\ref{director-full}) become
\begin{subequations}
    \begin{eqnarray}
           \frac{D(\rho\bm{ n})}{Dt}& =&(\rho\bm{I} \bm{n}-\bm{T}):(\beta \bm{E}+\bm{W})-\frac{\rho\bm{n}}{\Pe_r},\label{n0_eqn}
\\
    \frac{D(\rho\bm{Q})}{Dt}&=&           \beta\rho\left(\bm{E}\cdot(\bm{Q}+\bm{I}/2)+(\bm{Q}+\bm{I}/2)\cdot\bm{E}\right)+\rho(\bm{W}\cdot\bm{Q}-\bm{Q}\cdot\bm{W})-2\beta\bm{G}: \bm{E}-\frac{4\rho\bm{Q}}{\Pe_r}.\label{q0_eqn}
\end{eqnarray}\label{outer_eqns}
\end{subequations}
Because bacterial density is constant along the streamlines here, we seek solutions to Eqs.\,(\ref{outer_eqns}) that are independent of spatial position and time. This means setting the sum of all source terms on the right-hand-sides of Eqs.\,(\ref{outer_eqns}) to zero, and solving the resulting algebraic equations. This task is simplified in shear flow as the fluid tensors have the following forms:
\begin{align}
     \bm{E}=\frac{1}{2}\begin{pmatrix}
       0&1
       \\1 & 0
   \end{pmatrix}, \quad     \bm{W}=\frac{1}{2}\begin{pmatrix}
       0&1
       \\-1 & 0
   \end{pmatrix} .
\end{align}
The solution of \eqref{n0_eqn} which is independent of space and time, is $\bm{n}=\bm{0}$, namely that there is no biased swimming direction far from the surface.  Similarly, solving \eqref{q0_eqn} we find the following leading-order solution that describes a balance between the Jeffery-orbit distribution, \cite{jeffery1922motion}, and rotational diffusion:
\begin{align}
   \bm{Q} \sim \frac{\beta \Pe_r^2}{4(16+\Pe_r^2)}\begin{pmatrix}
       \Pe_r & 4
       \\ 4 & -\Pe_r
   \end{pmatrix}.\label{outer_Q}
\end{align}
For circular bacteria ($\beta=0$), the nematic order tensor in \eqref{outer_Q} is equal to zero, as the angular distribution of circular bacteria is uniform. For elongated bacteria ($\beta>0$), the first component of the tensor is positive, reflecting that the bacteria become preferentially aligned parallel to the flow in this case, as predicted by the classical Jeffery-orbit distribution. Note that it is consistent for the bacteria to be preferentially aligned parallel to the flow 
according to \eqref{outer_Q} but for there to be no biased swimming direction, $\bm{n}=\bm{0}$, as the bacteria are equally likely to be facing upstream or downstream and therefore net movement from swimming is zero. \eqref{outer_Q} and the solution $\bm{n}=\bm{0}$ provide self-consistent inlet boundary conditions for $\bm{Q}$ and $\bm{n}$ (negating the need for an inlet boundary layer), which automatically satisfy Eqs.\,(\ref{outer_eqns}) throughout the region away from the surface. 
\subsubsection*{Active Lévêque boundary layer}\label{bdy-layer}

\begin{figure}[h]
    \centering
    \includegraphics[width=0.85\linewidth]{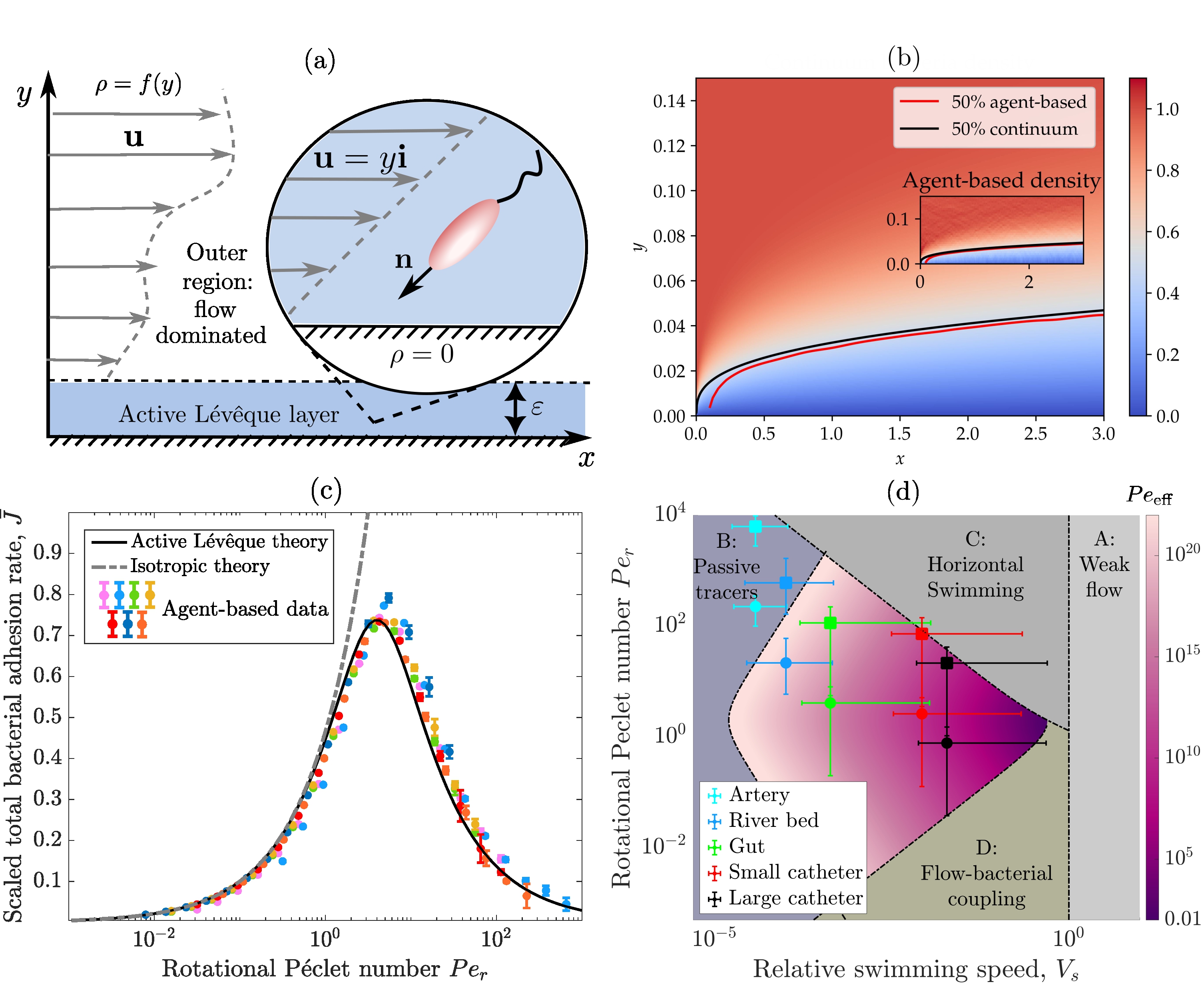}
       \caption{Continuum model for bacterial adhesion in flow. (a) Continuum model framework; far from the surface, bacteria are transported by a general laminar flow; in the thin active Lévêque layer the flow is well approximated by shear flow and bacteria density is depleted. (b) An analytical solution of the active Lévêque problem for $\Pe_r=1$, $V_s=0.01$, $\beta=0$. Inset: bacterial density from agent-based simulations with matching contours. Both plots show 50\% density contours demonstrating agreement between agent-based density and continuum model.   (c) Scaled total bacterial adhesion $\bar{J}$ (defined in \eqref{J-bar}), is shown as $\Pe_r$ varies. Agent-based data with colours matching Fig.\,\ref{fig:agent}b is shown with the standard deviation showing variation over individual simulations. Solid and dashed lines show predicted  adhesion rates from our active Lévêque theory (using \eqref{J-dimless} with $\Pe_{\text{eff}}$) and isotropic theory (using \eqref{J-dimless} with $\Pe_{\text{quies}}$), with no fitting parameters required. (d) Value of effective P\'eclet number $\Pe_{\text{eff}}$ (colourbar) is shown for varying $\Pe_r$ and $V_s$ (colourbar). Markers: \textit{E. coli} - circles, \textit{P. aeruginosa} - squares. The assumptions required to calculate our adhesion rate are satisfied for applications with data points in central coloured region. Regions A-D define regimes where our analysis does not formally hold - see \S\ref{adhesion-result}. }
    \label{fig:regimes}
\end{figure}

To predict the adhesion rate of the bacteria in fluid flow we now consider the behaviour of the suspension near the surface. The singular nature of the continuum system we derive means that the behaviour of the suspension near the surface is significantly different to the behaviour far from the surface. This boundary layer effect arises because adhesion significantly alters the suspension density in a thin region of height $\eps \ll 1$ near the surface, where we determine $\eps$ in terms of the system parameters. Our approach generalises the classic Lévêque theory for passive particles, in which thermal diffusion drives spatial gradients, which in turn lead to adhesion. Here, by contrast, bacterial motility leads to spatial gradients that facilitate adhesion. Hence, we must track the bacterial swimming direction or, equivalently, the bacterial orientation. We define a boundary layer coordinate $\tilde{y}=y/\eps = O(1)$ and denote variables in this region with tildes. A schematic of the boundary layer structure is given in Fig.\,\ref{fig:regimes}a.

We first note that the steep vertical spatial gradients in the boundary layer mean that the conservation equation \eqref{rho_eqn} for bacterial density $\tilde \rho$ becomes a balance between horizontal transport by the flow and vertical swimming, yielding: 
\begin{align}
  \tilde{y}\pdif{\tilde{\rho}}x+\frac{V_s}{\eps^2}\pdif{(\tilde{\rho} \tilde{n}_{y})}{\tilde{y}}&=0.\label{pre-Lévêque}
  \end{align}
In \eqref{pre-Lévêque}, the vertical component of the mean orientation of the bacteria $\tilde n_y$ determines the amount of transport towards the surface. Our analysis below will allow us to write the mean orientation vector $\bm{\tilde n}$ in terms of $\tilde \rho$ and the fixed system parameters, which will allow us to derive a closed PDE for $\tilde \rho$ from \eqref{pre-Lévêque}. 

We show in the Supplementary Information (\S \ref{S3}) that a small anisotropy arises in the mean orientation field due to a balance between
(i) shear alignment, (ii) rotational diffusion, and (iii) swimming of bacteria down vertical density gradients. Given this, we analytically determine the following leading-order solution for the mean orientation of the bacteria, $\bm{\tilde{n}}=(\tilde{n}_{x}, \tilde{n}_{y})$:
\begin{align}
\tilde{\rho} \tilde{n}_{y}\sim-\frac{ 4\Pe_r V_s(32+(2 - \beta)(1 - \beta)\Pe_r^2)}{\eps(16+\Pe_r^2)(16+(4-\beta^2)\Pe_r^2)}\pdif{\tilde{\rho}}{\tilde{y}}, \quad \tilde{\rho} \tilde{n}_{x}\sim-\frac{\Pe_r^2 V_s(64+46\beta+(4-\beta^2)\Pe_r^2)}{\eps(16+\Pe_r^2)(16+(4-\beta^2)\Pe_r^2)}\pdif{\tilde{\rho}}{\tilde{y}}.\label{polar-bdy} 
\end{align}
The small anisotropy generated within the boundary layer is sufficient to produce enough vertical bacterial transport towards the wall to balance the horizontal transport by the flow in \eqref{pre-Lévêque}. Because the anisotropy in the mean orientation field is small, the nematic order tensor remains equal to \eqref{outer_Q} and therefore spatially uniform at leading-order. 
Using the vertical component $\tilde{n}_y$ of \eqref{polar-bdy} in \eqref{pre-Lévêque}, we deduce that
  \begin{align}
\tilde{y}\pdif{\tilde{\rho}}x-\frac{1}{\eps^3\Pe_{\text{eff}}}\spdif{\tilde{\rho}}{\tilde{y}}&=0, \qquad \text{where } \Pe_{\text{eff}}=\frac{ (16+\Pe_r^2)(16+(4-\beta^2)\Pe_r^2)}{4\Pe_rV_s^2(32+(2-\beta)(1-\beta)\Pe_r^2)}, \label{Lévêque}
\end{align}
which is a key result of our analysis: a closed PDE that describes the bacterial density near the surface. Importantly, \eqref{Lévêque} has the same functional form as the classic Lévêque boundary layer problem, however we now capture the effect of bacterial motility and flow on dispersion near the surface through the effective P\'eclet number, $\Pe_{\text{eff}}$. Seeking a self-consistent balance in \eqref{Lévêque} gives an explicit prescription of the boundary layer thickness:
\begin{align}
\eps \tilde{y} \sim (x/\Pe_{\text{eff}})^{1/3}. \label{BL thickness}
\end{align}
As reassuring consistency checks for the effective Peclet number, \eqref{Lévêque},  we note that in the limit $\Pe_r\to0$, corresponding to weak flow or strong rotational diffusion, $\Pe_{\text{eff}} \sim 2/V_s^2\Pe_r=\mathcal{U}\mathcal{L}/D_{\text{quies}}=\Pe_{\text{quies}}$ which is the classic P\'eclet number for bacteria when dispersion is controlled by motility alone \citep{berg1993random}. Additionally, for spherical bacteria ($\beta=0$) the effective P\'eclet number $\Pe_{\text{eff}}=(4+\Pe_r^2)/2\Pe_rV_s^2,$ which is the 2D equivalent of the $D_{yy}$ component of the diffusion tensor calculated in \cite{bearon2003extension}.

\subsection{Bacterial adhesion rates}\label{adhesion-result}

The active Lévêque problem \eqref{Lévêque} (accompanied by appropriate boundary and matching conditions) admits a classic similarity solution. This solution is a uniformly accurate description of the density both in the boundary layer and far from the surface
\begin{align}
     \rho(x,y)=\frac{1}{\Gamma(1/3)}\Gamma\left(\frac{1}{3},\frac{\Pe_\text{eff} \,y^3}{9x}\right), \label{solution}
 \end{align} which we write in terms of the standard $(x,y)$ coordinates. In \eqref{solution}, $\Gamma(a,z)$ is the lower incomplete gamma function and $\Gamma(a)$ is the gamma function. The analytic solution \eqref{solution} obtained from our boundary layer analysis shows excellent quantitative agreement with the bacteria density calculated from our full agent-based simulations (Fig.\,\ref{fig:regimes}b) everywhere except near $x,y=0$ (as expected) where there is an additional boundary layer. 
 Using \eqref{solution}, we calculate a key result of our analysis: the bacterial adhesion rate to the surface:\begin{align}
 J(x)=\frac{1}{\Pe_{\text{eff}}}\left.\pdif{\rho}{{y}}\right|_{\tilde{y}=0}=\frac{3^{1/3}}{\Gamma(1/3)\Pe_{\text{eff}}^{2/3}x^{1/3}}.\label{J-dimless}
 \end{align}
 Given the mathematical equivalence noted above, \eqref{J-dimless} shares the same functional form as
 the classic result derived for passive particles in \cite{levêque1928lois}, but here with an effective P\'eclet number that we have calculated explicitly, which captures the interacting effects of bacterial motility, shear-induced alignment and bacterial shape on adhesion. 
In dimensional form, this adhesion rate is
  \begin{align}
\hat{J}(\hat{x})=\frac{3^{1/3}\rho_\infty D_{\text{eff}}^{2/3}\dot\gamma^{1/3}}{\Gamma(1/3)\hat{x}^{1/3}}, \quad \quad
D_{\text{eff}}=\frac{4D_rV_s^2(32D_r^2+(2-\beta)(1-\beta)\dot\gamma^2)}{(16 D_r^2 + (4 
-\beta^2) \dot\gamma^{2})(16D_r^2+\dot\gamma^2)},\label{dim-adhesion}
 \end{align}
 where $D_{\text{eff}}$ represents the effective diffusion in the boundary layer, $\hat{x}$ is the dimensional distance from the inlet and $\rho_\infty$ is the bacterial density at the inlet. Adhesion decreases with distance from the inlet at a rate of  $\hat x^{-1/3}$, and increases with $D_{\text{eff}}^{2/3}$. 
 Elongated bacteria ($\beta>0$) have a smaller effective diffusivity and therefore adhere less than spherical bacteria ($\beta=0$). We interpret this physically as elongated bacteria preferentially orientating parallel to the flow, reducing the number of bacteria orientated towards the boundary. The reduction in adhesion rate of elongated bacteria is most pronounced at high shear rates because rotational diffusion is unable to significantly reorient bacteria towards the wall.

A notable feature of $D_{\text{eff}}$ in \eqref{dim-adhesion} is that it depends on the fluid shear rate, so that the bacterial adhesion rate $\hat{J}$ depends non-monotonically on the fluid shear rate, with maximal adhesion occurring at a critical shear rate  $\dot{\gamma}_{\text{crit}}$. In general, $\dot{\gamma}_{\text{crit}}$ depends on both rotational diffusivity $D_r$ and the shape parameter $\beta$. For spherical bacteria, we obtain the analytic result $\dot{\gamma}_{\text{crit}}=2D_r/\sqrt{3}$. Using parameter values from \Cref{Table:dimensional}, we calculate that maximum adhesion occurs at $\dot{\gamma}_{\text{crit}}=1.1 s^{-1}$ for \textit{E. coli} and $\dot{\gamma}_{\text{crit}}=0.039s^{-1}$ for \textit{P. aeruginosa}. The non-monotonic dependence of the adhesion rate on fluid shear rate explains the agent-based findings shown in Fig.\,\ref{fig:agent}b: our calculated adhesion rate scales like $\hat{J}\sim\dot{\gamma}^{1/3}$ for smaller shear rates until $\dot\gamma\approx\dot{\gamma}_{\text{crit}}$, then scales with $\hat{J}\sim\dot{\gamma}^{-1}$.
Moreover, our analysis successfully predicts a specific scaling collapse of the total scaled adhesion rate $\bar{J}$, defined in \eqref{J-bar}, for all agent-based simulations (Fig.\,\ref{fig:regimes}c). The adhesion predicted by our calculated effective diffusivity $D_{\text{eff}}$ correctly shows the maximum adhesion and the decrease in adhesion at high shear (high $\Pe_r$), in contrast to the adhesion calculated using the isotropic quiescent diffusivity $D_{\text{quies}}=V_s^2/(2D_r)$ (grey dashed line Fig.\,\ref{fig:regimes}c). While the quiescent approximation works well for low shear (low $\Pe_r$) it does not capture the non-monotonic nature of the adhesion as the shear rate increases. 

The non-monotonic adhesion rate can be explained physically by examining the orientation and mean swimming speed of the bacteria in the boundary layer, given in \eqref{polar-bdy}.
For small $\dot{\gamma}$ (small $\Pe_r$), diffusive reorientation is comparatively rapid, and bacteria quickly lose any information about orientational bias arising from density gradients. Therefore, mean vertical swimming is negligibly small, and adhesion is minimal. 
At moderate shear $\dot{\gamma}$ (moderate $\Pe_r$), overall swimming persistence is stronger which increases mean swimming speed, and correspondingly, the adhesion rate increases. However, at high shear (large $\Pe_r$), the increased flow rate rotates the bacteria back upstream,  minimising the total available to swim into the boundary and reducing adhesion.

The adhesion rate derived in this paper, \eqref{dim-adhesion}, is valid for the typical parameters associated with a wide range of industrial and medical scenarios in which bacterial adhesion and subsequent biofilm formation are problematic. 
The shear rate near surfaces, also called the wall shear rate, and the flow lengthscales for several such scenarios are presented in \Cref{Table:flow}. 
The motility parameters for common pathogens \textit{E. coli} and \textit{P. aeruginosa} in these scenarios are presented in \Cref{Table:dimensional}. The value of the dimensionless effective P\'eclet number as a function of dimensionless parameters ($\Pe_r$, $V_s$) is shown in Fig.\,\ref{fig:regimes}d for $\beta=0$.
Specific examples of the pathogens \textit{E. coli} and \textit{P. aeruginosa} in these industrial and medical settings are shown through the overlayed datapoints in Fig.\,\ref{fig:regimes}d.  For the majority of these applications, the data fall within the coloured region in which our formal analysis is valid, and therefore the system parameters satisfy the requirements to apply our result \eqref{dim-adhesion} to predict adhesion. Furthermore, for all scenarios except \textit{E. coli} adhesion in large catheters, the wall shear rate is greater than the critical shear rate at which maximum adhesion occurs, $\dot\gamma>\dot \gamma_{\text{crit}}$. Our theory therefore predicts that shear-induced reorientation of bacteria reduces adhesion in the industrial and medical scenarios considered and that a prediction based on the classic diffusion coefficient $D_{\text{quies}}$ would overestimate adhesion in these scenarios. 

Our asymptotic theory relies on specific parameter restrictions, which will not hold everywhere in  parameter space. We label the regions in which the restrictions do not hold as A--D in Fig.\,\ref{fig:regimes}b, and briefly summarise what happens in each of these regions. In A, the bacterial swimming speed is comparable to the flow speed, and therefore swimming alters the bacterial density far from the surface, modifying Eqs.\,(\ref{rho0_eqn})--(\ref{outer_eqns}). In B, the boundary layer thickness \eqref{BL thickness} is less than the body length of a single bacterium and, therefore, bacteria are transported like ballistic passive tracer particles. In C, shear-induced reorientation is dominant and \eqref{polar-bdy} does not well describe the bacterial orientation. In this regime, the majority of bacteria are oriented parallel to the surface and there are no longer sufficient numbers of bacteria facing the boundary for our model to accurately capture adhesion. In D, swimming is very strong relative to rotational diffusion, so the nematic order tensor is not spatially uniform and active flows are generated, modifying our result for mean orientation \eqref{polar-bdy}. Regions A-D are defined in terms of the dimensionless parameters $(V_s,\Pe_r)$ and the boundary layer thickness $\eps$ as follows A: $V_s\geq1$, B:
$\eps=5\times10^{-4}$ (calculated for a bacterium body length of 5$\mu$m and a flow lengthscale $\mathcal L=$\,1cm), C: $\Pe_r\geq \eps^{-1}$ and D: $V_s\leq \eps$.
To determine bacterial adhesion in regions A and D, further analysis is required. In regions B and C, bacteria have less opportunity to swim into the boundary, suggesting that the adhesion rate \eqref{dim-adhesion} can provide an upper bound for adhesion even in these regions, thus covering a wider range of relevant flow rates and bacteria species than the formal requirements might suggest.

\section{Discussion and conclusion}\label{Discussion}
Using a combination of agent-based and continuum modelling, we have provided a quantitative upper bound for the adhesion rate of bacteria to surfaces, specifically isolating the effects of bacterial motility, bacterial shape and fluid flow. Our agent-based simulations demonstrate that there is a maximum in adhesion at an intermediate shear rate. Our continuum model and analysis demonstrate that capturing this maximal adhesion rate requires accounting for non-isotropic diffusion, which arises from shear-induced bacterial reorientation. We achieve this by systematically deriving an accurate dispersivity that applies across a range of shear rates. We have found that the classical quiescent dispersivity $D_{\text{quies}}=\mathcal{V}_s^2/2D_r$ can only account for adhesion at low flow rates. Our new, effective dispersivity applies at shear rates relevant to a wide range of industrial applications. 

We have integrated previous bacterial dispersivity results into an active Lévêque boundary layer theory to predict surface adhesion. Our predicted maximal adhesion rate occurs when axial diffusion is minimal, which corresponds to intermediate shear rates. Such a reduction in axial diffusion has been found in studies of bacterial suspensions in non-adhesive pipes \citep{peng2020upstream,chilukuri2015dispersion,jiang2019dispersion}.
A reduction in adhesion at high flow rates was also observed in simulations of surface adhesion of spherical bacteria in the presence of a chemotactic gradient \cite{bearon2003extension}. In the absence of surface adhesion, motility and shear-induced reorientation have been found to produce active phenomena that differentiate bacterial suspensions from passive mixtures. For example, reorientation has been found to lead to persistent upstream swimming close to non-adhesive surfaces \cite{kaya2012direct, Ezhilan2015}. 
The interaction between these effects in settings with non-uniform shear, such as in channel flows, has also been shown to produce wall accumulation as the density near the centre of the pipe is depleted \cite{Ezhilan2015,rusconi2014bacterial}. We have demonstrated that both bacterial motility and shear-induced reorientation can drive significant changes in bacterial adhesion, especially in high-flow regimes.

 Our work contributes to a general understanding of the complex process of bacterial surface adhesion and colonisation. We have isolated the effects of bacterial motility, fluid transport and shear-induced cell reorientation on adhesion to perfectly absorbing surfaces. To disentangle the specific contribution of these effects, we have not included several further biophysical processes that have been found in experiments to affect bacterial adhesion, such as reversible adhesion, surface chemistry and cell appendages \cite{palalay2023shear,moreira2014effects}. Our prediction will enable us to isolate the effects of such additional complex processes by comparing measured adhesion rates to our predicted maximal rate. Overall, we have demonstrated how fluid flow and motility can both facilitate and limit attachment in the fundamental problem of bacterial surface adhesion.

\subsection*{Acknowledgments and Funding}
E.F.Y is supported by an EPSRC National Fellowship in Fluid Dynamics [EP/X027902/1]. P.P. is supported by a UKRI Future Leaders Fellowship [MR/V022385/1]. 

 \subsection*{Data, Materials, and Software Availability}  The computational tools for agent-based model, data processing scripts and analytical calculation notebooks are
freely available at https://github.com/Edwina-Yeo/bacterial-surface-adhesion.
\subsection*{Supporting Material}
Full mathematical analysis of continuum model, numerical calculation  of inlet distribution for agent based simulations are given in supplementary information appendix \ref{supplementary}.

\bibliography{lib}

\appendix

\section{Agent-based numerical method}\label{agent-num}We simulate the dynamics of bacteria independently in a numerical domain $x\in[0,3]$, $y\in[0,H]$ that is large enough to contain the diffusive boundary layer. We solve Eqs.\,(\ref{SDE-vel}) and (\ref{SDE-angle}) in dimensionless form using the explicit stochastic time-stepping scheme Euler-Maruyama with a fixed timestep of $\delta t=10^{-2}$. At each timestep, 100 bacteria enter the domain at $x=0,y=y_i$, where the inlet positions $y_i$ are drawn from a probability distribution that describes a uniform suspension advected by shear flow: $p(y_i)=H\sqrt{y_i}$. 
The initial inlet orientations are sampled from the numerically calculated diffusive Jeffery-orbit distribution \cite{talbot2024exploring} (see Supplementary Information \S \ref{S4}).

For the spatial density comparison in Fig.\,\ref{fig:regimes}b, we solve the system until $t=1000$ using a domain height of $H=0.8$.  We record the location of each bacteria over time and form a histogram in $(x,y)$ by calculating the average number of bacteria in each bin over time.  We discard the data for $t<300$ to ensure the boundary layer has reached steady state. We normalise the density field according to the expected number of bacteria in each box (in the absence of motility), as determined by the domain size, the number of boxes and the number of bacteria entering the domain per $\delta t$.  
To measure bacterial adhesion rate across a wide range of motility parameters, we solve the system until $t=10300$ with a domain height of $H=1.5$. We record the locations of any adhesion events with the boundary that occur within the time interval $[300,10300]$. We split this into time intervals of length 2500 which, since the bacteria are decoupled, is equivalent to carrying out 4 independent simulations over a time window of 2500. We then calculate the total adhesion occurring over $x\in[0.5,3]$, and average over each time window to obtain the mean adhesion rate $J_{\text{num}}$ and the standard deviation around that mean. We discard the region near $x=0$ as our boundary layer solution does not hold here. To allow comparison between dimensionless agent-based simulations, the total adhesion rate plotted in Fig.\,\ref{fig:agent}b is in dimensional form as $J_{\text{num}}\dot{\gamma}$, which has units of cells/s. 
The scaled total adhesion rate, $\bar{J}$, plotted in Fig.\,\ref{fig:regimes}c is defined as the integral of the dimensionless adhesion rate $J$ (\eqref{J-dimless}) along the surface, scaled by our predicted dependence of the adhesion rate on swimming speed and shape, $S(\beta,\Pe_r,V_s)$. We calculate $\bar{J}$ from the mean total agent-based adhesion rate $J_{\text{num}}$ as follows:
 \begin{align}
 \bar{J}=S(\beta,\Pe_r,V_s)\int_{0.5}^3J\,\mathrm{d}x=\frac{J_{\text{num}}S(\beta,Pe_r,V_s)}{\rho_{\infty}},\quad S(\beta,\Pe_r,V_s)=\frac{(16+(4-\beta^2)\Pe_r^2)^{2/3}}{V_s^{4/3}(32+(2-\beta)(1-\beta)\Pe_r^2)^{2/3}}\label{J-bar},
 \end{align} where the value $\rho_\infty$ converts the agent-based adhesion rate, which has units number of bacteria per unit time, to the dimensionless adhesion rate. This is calculated by equating the number of particles arriving per $\delta t$ in the agent-based simulations with the continuum inlet flux, giving $\rho_\infty= 8\times10^3$. In \eqref{J-bar}, $V_s$ and $\Pe_r$ are calculated from dimensional motility parameters using the shear rate $\dot{\gamma}$ and a flow lengthscale of $\mathcal{L}=750\,\mu $m.

\section{Flow and bacteria parameters}
The shear rate (also known as the wall shear rate) and the flow lengthscales for industrial and medical scenarios are presented in \Cref{Table:flow}. Dimensional bacterial parameter values for \textit{E. coli} and \textit{P. aeruginosa} are presented in \Cref{Table:dimensional}. These two parameter sets are used to calculate the data points in Fig.\,\ref{fig:regimes}d. We approximate \textit{E. coli}'s run-and-tumble reorientation dynamics as continuous reorientation via rotational diffusion by replacing the rotational diffusion coefficient with the tumble rate. This approximation is valid when the bacteria are dilute and are not subject to external forces \cite{cates2013active}.
\begin{table}[h]

\centering

\caption{Wall shear rates and flow lengthscales of industrial and medical systems prone to biofilm formation.}\label{Table:flow}

\begin{tabular}{llll}

Scenario              & WSR (1/s) & Lengthscale & Source \\ 
\midrule
Small Catheter        & 0.12-5.09                           & 1 mm                       &  \cite{bull2025different,lee2024analysis}      \\
Large Catheter        & 0.038-1.51                          & 1.5 mm                     &    \cite{bull2025different,lee2024analysis}    \\
Human small intestine & 0.2-80                              & 1.25 cm                    &     \cite{lindner2021physiological,helander2014surface}   \\
Slowly flowing river  & 5.9-21.7                            & 1 cm                       &  \cite{droppo2011modelling}      \\
Human coronary artery & 100-350                             & 2.6 mm                    &    \cite{doriot2000vivo,dodge1992lumen}    \\
\bottomrule
\end{tabular}

\end{table}

\begin{table}[h]

\centering
\caption{Bacterial motility parameter values.}\label{Table:dimensional}.    
\begin{tabular}{llllll}
Name                 & Strain   & Symbol          & Units                       & Value
& Source                            \\     \midrule            
Swimming speed  \textit{E. coli}   &  HCB437   & $\mathcal{V}_s$ & $\mu$m/s                    & $22\pm5$                    & \cite{drescher2011fluid}    \\
Swimming speed   \textit{P. aeruginosa}  &   PAO1   & $\mathcal{V}_s$ & $\mu$m/s                    & $23\pm6$                   &\cite{khong2021dynamic} \\
Rotational diffusion coeff. \textit{E. coli} & RP437 & $D_r$           & 1/s                   &   1                        &        \cite{saragosti2012modeling}\\
Rotational diffusion coeff. \textit{P. aeruginosa} & PAO1   & $D_r$           & 1/s                   &   0.036                        &  \cite{cai2016singly}      \\
Bretherton param. \textit{E. coli} & HCB437  & $\beta$         & -                           &  0.88                & \cite{drescher2011fluid}        \\
Bretherton param.  \textit{P. aeruginosa} &  Species averages given   & $\beta$         & -                           & 0.88               &   \cite{diggle2020microbe}    \\
                  \bottomrule
\end{tabular}

\end{table}

\newpage
\section{Supplementary Information}\label{supplementary}
\subsection{Derivation and closure of the continuum model}\label{S1}
In this section we derive Eqs.\,(3-5)  from the agent-based model defined in Eqs.\,(1-2).
We use the model for the transport of a collection of bacteria at the continuum scale following the orientational moment framework from \citep{saintillan2015theory}, combined with a 2D model closure from \citep{theillard2019computational,fylling2024multi} to define the higher-order tensors, Eqs.\,(6).
The model is nondimensionalised using a reference velocity scale $\mathcal{U}=\dot{\gamma}\mathcal{L}$ for the external flow, with reference lengthscale $\mathcal{L}$ that is much larger than the size of an individual bacteria, and the timescale of fluid transport $\dot{\gamma}^{-1}$. We start by considering the 
probability density function $\psi(\bm{x},\bm{s},t)$, which describes the probability of finding a bacteria at position $\bm{x}$ with orientation $\bm{s}$ at time $t$. This probability density $\psi$ evolves according to the following (dimensionless) Fokker-Planck equation
      \begin{align}
              \pdif{\psi}t&+ \na_{\bm{x}} \cdot \left(\bm{v} \psi\right)
        +\hat{\na}_{\bm{s}}\cdot \left( \bm{v}_{s}\psi\right)-\frac{1}{Pe_r}\hat{\na}^2_{\bm{s}}\psi
       =0,\label{FP}
\end{align}
where  ${\na}_{\bm{x}}$ is the gradient operator with respect to spatial variables, $\hat{\na}_{\bm{s}}=(\bm{I}-\bm{s}\bm{s})\na_{\bm{s}}$ is the gradient operator on the unit sphere where $\bm{I}$ is the identity tensor, $\bm{s}\bm{s}$ defines a rank-2 tensor with components [$\bm{s}\bm{s}]_{ij}=s_is_j$ and $\na_{\bm{s}}$ is the gradient operator with respect to orientational variables. The Laplacian operator on the surface of the unit sphere is defined $\hat{\nabla}^2_{\bm{s}}=\hat{\nabla}_{\bm{s}}\cdot\hat{\nabla}_{\bm{s}}$.
From left to right the terms in \eqref{FP} are: the evolution of probability in time, the spatial advection of probability, the angular advection of probability and rotational diffusion.
The velocity vectors in \eqref{FP} are the dimensionless individual bacterial velocity and angular velocity: $\bm{v}=\bm{u} +V_s\bm{s}$ and $
\bm{v}_{s}=(\beta\bm{E}+\bm{W})\bm{s}$ with rate of rotation $\bm{W}$ and strain tensors $\bm{E}$ defined using the dimensionless velocity field: 
\begin{align}
    \bm{W}=\frac{1}{2}(\na\bm{u}-\na{\bm{u}}^T), \quad     \bm{E}=\frac{1}{2}(\na\bm{u}+\na{\bm{u}}^T).
\end{align}
As described in the main text the model has two dimensionless parameters: $V_s=\mathcal{V}_s/\dot{\gamma}\mathcal{L}$, which is the ratio of bacterial swimming speed to the flow speed and $Pe_r=\dot{\gamma}/D_r$, the rotational P\'eclet number, which is the ratio of fluid rotation to rotational diffusion. 
We define the density $\rho$, the mean bacterial orientation $\bm{n}$ (also known as the polar order parameter)
and the nematic order tensor $\bm{Q}$ of the collection of bacteria as the zeroth, first and second angular moments of the probability $\psi$ as follows
\begin{align}
  \rho(\bm{x},t)=  \int\!\psi \, \mathrm{d}\bm{s},\ \quad
    \bm{n}(\bm{x},t)= \frac{1}{\rho}\int\!\bm{s}\psi\,\mathrm{d}\bm{s},\ \quad   \bm{Q}(\bm{x},t)=\frac{1}{\rho} \int\!\left(\bm{s}\bm{s}-\frac{\bm{I}}{2}\right)
\psi\, \mathrm{d}\bm{s}.\label{definition}
\end{align}
The evolution of the density $\rho$, mean orientation $\bm{n}$ and nematic order tensor $\bm{Q}$ can be found by taking the zeroth, first and second moments of \eqref{FP}, which gives the following system, as presented in \cite{saintillan2015theory,theillard2019computational}:
\begin{subequations}
\begin{align}    
     \frac{D\rho}{Dt}=&-V_s\na\cdot(\rho\bm{n}),\label{rho_eqnS}\\
           \frac{D(\rho\bm{ n})}{Dt}=&-V_s\left(\na\cdot(\rho\bm{Q})+\frac{1}{2}\na\rho\right) +(\rho\bm{I} \bm{n}-\bm{T}):(\beta \bm{E}+\bm{W})-\frac{\rho\bm{n}}{Pe_r},\label{n_eqnS}
\\
    \frac{D(\rho\bm{Q})}{Dt}=&-V_s\left(\na\cdot\bm{T}+\frac{\bm{I}}{2}\na\cdot(\rho\bm{n})\right)+
           \beta\rho\left(\bm{E}(\bm{Q}+\bm{I}/2)+(\bm{Q}+\bm{I}/2)\bm{E}\right)+\rho(\bm{W}\bm{Q}-\bm{Q}\bm{W})-2\beta\bm{G}: \bm{E}-\frac{4\rho\bm{Q}}{Pe_r}.\label{q_eqnS}
           \end{align}\label{full-eqns}
\end{subequations}
The higher-order moments of $\bm{s}$, the orientation vector, are: $\bm{T}$, the third moment and $\bm{G}$, the fourth moment. These are defined as
\begin{align}
    \bm{T}=\int\!\bm{s}\bm{s}\bm{s}
\psi\, \mathrm{d}\bm{s},\quad\quad     \bm{G}=\int\!\bm{s}\bm{s}\bm{s}\bm{s}
\psi\, \mathrm{d}\bm{s}.
\end{align}
In this paper we follow \citep{theillard2019computational} and use a truncated angular harmonic closure to define $\bm{T}$ and $\bm{G}$. In 2D, this is defined as follows
\begin{align}
    \psi(\bm{x},\bm{s},t)=\frac{\rho}{2\pi}\left(1+2\bm{s}\cdot\bm{n}+4\bm{s}\bm{s}:\bm{Q}\right),\label{harmonic_closure}
\end{align}
and is equivalent to approximating the probability $\psi $ using a truncated Fourier series \citep{theillard2019computational}. Consequentially, \eqref{harmonic_closure} becomes a better approximation the less important higher-order Fourier modes are for $\psi$, for example if $\psi$ is close to a uniform distribution. We quantify the accuracy of \eqref{harmonic_closure} in \S\ref{validity}.
Under this closure assumption we have the following definitions of the requisite higher-order moments, as stated in \citep{fylling2024multi}:
\begin{eqnarray}
\bm{T}_{ijk}&=&\frac{\rho}{4}(\delta_{ij}n_{k}+\delta_{ik}n_{j}+\delta_{jk}n_i),\label{def_pppS}\\
\bm{G}_{ijkl}&=&\frac{\rho}{8}(\delta_{ij}\delta_{lk}+\delta_{ik}\delta_{jl}+\delta_{il}\delta_{jk})+\frac{\rho}{6}(\delta_{ij}Q_{lk}+\delta_{ik}Q_{jl}+\delta_{il}Q_{jk}+\delta_{jk}Q_{il}+\delta_{jl}Q_{ik}+\delta_{jk}Q_{ij}).\label{def_ppppS}\end{eqnarray} 
\subsection{Derivation of solution far from the surface}\label{outer-section}
In this section we derive the analytical solutions for the mean orientation vector $\bm{n}$ and nematic order tensor $\bm{Q}$ that hold far from the surface. The leading-order equations, which hold far from the surface, are obtained from Eqs.\,(\ref{full-eqns}) by removing terms including $V_s$:
\begin{subequations}
    \begin{eqnarray}
 \frac{D\rho}{Dt}&=0,\label{rho0_eqnS}\\
           \frac{D(\rho\bm{ n})}{Dt}& =&(\rho\bm{I} \bm{n}-\bm{T}):(\beta \bm{E}+\bm{W})-\frac{\rho\bm{n}}{Pe_r},\label{n0_eqnS}
\\
    \frac{D(\rho\bm{Q})}{Dt}&=&           \beta\rho\left(\bm{E}(\bm{Q}+\bm{I}/2)+(\bm{Q}+\bm{I}/2)\bm{E}\right)+\rho(\bm{W}\bm{Q}-\bm{Q}\bm{W})-2\beta\bm{G}: \bm{E}-\frac{4\rho\bm{Q}}{Pe_r}.\label{q0_eqnS}
\end{eqnarray}
\end{subequations}
From \eqref{rho0_eqnS} we find that the bacterial density is constant with $\rho=1$ far from the surface, which motivates seeking solutions for the mean orientation vector $\bm{n}$ and nematic order tensor $\bm{Q}$ that are independent of space and time. Seeking a solution of \eqref{n0_eqnS} that is independent of space and time is equivalent to setting the sum of all the terms on the right-hand side equal to the zero vector: 
\begin{align}
    (\rho\bm{I} \bm{n}-\bm{T}):(\beta \bm{E}+\bm{W})-\frac{\rho\bm{n}}{Pe_r}=\bm{0}.
\end{align}
Using the definition of the fluid tensors in simple shear flow:
\begin{align}
     \bm{E}=\frac{1}{2}\begin{pmatrix}
       0&1
       \\1 & 0
   \end{pmatrix}, \quad     \bm{W}=\frac{1}{2}\begin{pmatrix}
       0&1
       \\-1 & 0
   \end{pmatrix} ,\label{fluid}
\end{align}
and after inserting the definition of $\bm{T}$, \eqref{def_pppS}, into \eqref{n0_eqnS} this gives two algebraic equations for the components of the mean orientation vector $\bm{n}=(n_x, n_y)$:
\begin{align}
        \left(\frac{1}{2}+\frac{\beta}{4}\right) \rho  n_y-\frac{\rho n_x}{\Pe_r}=0,\quad -\left(\frac{1}{2}-\frac{\beta}{4}\right) \rho  n_x-\frac{\rho n_y}{\Pe_r}=0.\label{n_components}
       \end{align}
 The only solution to this system for $\beta \in[0,1)$ is $(n_x, n_y)=(0,0)$. Therefore, away from the surface the bacteria have no biased swimming direction.  We now derive the nematic order solution, Eq.\,(9). We proceed in a similar manner, seeking a spatially uniform steady solution to Eq.\,\eqref{q0_eqnS}, equivalent to setting the sum of all terms on the right-hand-side equal to the zero tensor:
 \begin{align}
              \beta\rho\left(\bm{E}(\bm{Q}+\bm{I}/2)+(\bm{Q}+\bm{I}/2)\bm{E}\right)+\rho(\bm{W}\bm{Q}-\bm{Q}\bm{W})-2\beta\bm{G}: \bm{E}-\frac{4\rho\bm{Q}}{Pe_r}=\bm{0}\label{q_zero}
 \end{align}
The nematic order tensor is a symmetric rank 2 tensor with three unique components in 2D:
\begin{align}
    \bm{Q}=\begin{pmatrix}
        Q_{xx} & Q_{xy}\\Q_{xy} & Q_{yy}
    \end{pmatrix}.
\end{align}
 Hence, after inserting the fluid tensor solutions \eqref{fluid}, and the definition of $\bm{G}$ \eqref{def_ppppS}, into \eqref{q_zero} we have the following three algebraic equations for the three independent $\bm{Q}$-tensor components:
\begin{align}
 \rho Q_{xy}-\frac{4\rho Q_{xx}}{\Pe_r}=0,\quad
   \frac{\beta\rho}{4}+\frac{1}{2}\rho Q_{xx} (\beta-1)+\frac{1}{2}\rho Q_{yy} (\beta+1)-\frac{4\rho Q_{xy}}{\Pe_r}=0,\quad
-\rho Q_{xy}-\frac{4\rho Q_{yy}}{\Pe_r}=0.\label{q_components}
\end{align}
The solution of the system \eqref{q_components} is:\begin{align}
   \bm{Q}=\frac{\beta \Pe_r}{4(16+\Pe_r^2)}\begin{pmatrix}
       Pe_r & 4
       \\ 4 & -Pe_r
   \end{pmatrix}\label{outer_QS},
\end{align}
which describes the angular distribution of bacteria far from the surface. We note that $Q_{xx} > 0$ when $\beta > 0$ is consistent with the bacteria undergoing diffusive Jeffery-orbits. That is, elongated bacteria ($\beta > 0$) are more likely to be orientated parallel to the flow. We demonstrate the accuracy of \eqref{outer_QS} as $\beta$ and $\Pe_r$ vary in comparison to numerical solutions of the Fokker-Planck equation in the \S\ref{validity}. 

\subsection{Derivation of boundary layer solution}\label{S3}
In this section we derive the boundary layer solution for both the mean orientation $\bm{n}$, Eq.\,(11), and the density equation, Eq.\,(12), that hold close to the surface. 
To do this we carry out a boundary layer analysis on the full system, Eqs.\,(\ref{full-eqns}), identifying and examining a thin boundary layer of height $\eps \ll 1$ on the surface, where $\eps$ is to be determined in terms of the system parameters. We define a boundary layer co-ordinate $\tilde{y}=y/\eps = O(1)$ and denote inner variables with tildes. We then seek a steady solution exploiting $V_s\ll1$, that bacterial swimming is weak. The boundary layer transformation defines a rescaled gradient operator as follows
\begin{align}
    \tilde{\na}=\pdif{}x\bm{i}+\frac{1}{\eps}\pdif{}{\tilde y}\bm{j},\label{nabla}
\end{align}
where $\bm{i}$ and $\bm{j}$ are the unit vectors in the $x$- and $y$-directions, respectively. 
Before presenting the full boundary layer equations it is useful to state the values of the various flux terms on the right-hand-sides of Eqs.\,(\ref{full-eqns}) once they have been rescaled according to \eqref{nabla} and using the closure \eqref{harmonic_closure}:
\begin{align}
 \tilde{\na}\cdot(\tilde{\rho}\tilde{\bm{Q}})+\frac{1}{2}\tilde{\na}\tilde{\rho}&=\begin{pmatrix}
    \dfrac{\partial}{\partial x}(\tilde{\rho} \tilde{Q}_{xx})+ \dfrac{1}
    {\eps}\dfrac{\partial}{\partial \tilde y }(\tilde{\rho} \tilde{Q}_{xy})+\dfrac{1}{2}\dfrac{\partial\tilde \rho}{\partial x}\\ \dfrac{\partial}{\partial x}(\tilde{\rho} \tilde{Q}_{xy})+\dfrac{1}{\eps} \dfrac{\partial}{\partial \tilde y}(\tilde{\rho} \tilde{Q}_{yy})+\dfrac{1}{2\eps}\dfrac{\partial\tilde \rho}{\partial \tilde y}
 \end{pmatrix} \label{director-vel-bdy} ,\\
\tilde{\na}\cdot\bm{\tilde T}&=\begin{pmatrix}
    \dfrac{\partial \tilde T_{xxx}}{\partial x}+\dfrac{1}{\eps} \dfrac{\partial \tilde T_{yxx}}{\partial \tilde y} &       \dfrac{\partial \tilde T_{xxy}}{\partial x}+ \dfrac{1}{\eps}\dfrac{\partial \tilde T_{yxy}}{\partial \tilde y}\\
     \dfrac{\partial \tilde T_{xxy}}{\partial x}+\dfrac{1}{\eps} \dfrac{\partial \tilde T_{xyx}}{\partial \tilde y}&     \dfrac{\partial\tilde T_{xyy}}{\partial x}+ \dfrac{1}{\eps}\dfrac{\partial\tilde T_{yyy}}{\partial \tilde y}
\end{pmatrix}=\frac{1}{4}\begin{pmatrix}
    3\dfrac{\partial (\tilde \rho \tilde n_x)}{\partial x}+ \dfrac{1}{\eps}\dfrac{\partial(\tilde \rho \tilde n_y)}{\partial \tilde y} &   \dfrac{\partial (\tilde \rho \tilde n_y)}{\partial x}+ \dfrac{1}{\eps}\dfrac{\partial(\tilde \rho \tilde n_x)}{\partial \tilde y}\\\dfrac{\partial (\tilde \rho \tilde n_y)}{\partial x}+ \dfrac{1}{\eps}\dfrac{\partial(\tilde \rho \tilde n_x)}{\partial \tilde y}&   \dfrac{\partial (\tilde \rho \tilde n_x)}{\partial x}+ \dfrac{3}{\eps}\dfrac{\partial(\tilde \rho \tilde n_y)}{\partial \tilde y}
\end{pmatrix}\label{Q-bdy-T},
\\
\frac{\bm{I}}{2}\tilde{\na}\cdot(\tilde{\rho}\bm{n})&=\frac{1}{2} \left(
\dfrac{\partial (\tilde \rho \tilde n_x)}{\partial x}+ \dfrac{1}{\eps}\dfrac{\partial(\tilde \rho \tilde n_y)}{\partial \tilde y}\right)\begin{pmatrix}
   1 & 0\\0 &  1
\end{pmatrix}\label{Q-bdy-I}.\end{align}
The source terms in Eqs.\,(\ref{full-eqns}) which arise from rotation by the flow are unchanged in the boundary layer scaling as they do not include any spatial gradients. Therefore their values are equal to the algebraic expressions presented in the previous section: Eqs.\,(\ref{n_components}) and (\ref{q_components}).

We can now determine the rescaled steady boundary layer equations. First, the rescaled equation for density $\tilde\rho$, derived from \eqref{rho_eqnS} using \eqref{nabla} is
\begin{align}
 \eps {\tilde y}\pdif{\tilde\rho}x+\frac{V_s}{\eps}\pdif{(\tilde\rho \tilde n_y)}{\tilde y}+V_s\pdif{(\tilde\rho \tilde n_x)}x=0.\label{rhob_eqn}
\end{align}
First, examining \eqref{rhob_eqn} is it clear that the transport of bacteria is determined by the size of the two components of the mean orientation vector $\tilde n_x$ and $\tilde n_y$. 
The rescaled equations for the mean orientation vector components $\tilde n_x$ and $\tilde n_y$, derived from \eqref{n_eqnS} using Eqs.\,(\ref{nabla}), (\ref{director-vel-bdy}) and (\ref{n_components}) are as follows
\begin{subequations}
\begin{align}  
             \underbrace{\eps {\tilde y}\pdif{(\tilde\rho \tilde n_x)}x}_{(iv)}&=\underbrace{\left(\frac{1}{2}+\frac{\beta}{4}\right)\tilde \rho \tilde n_y}_{(i)}-\underbrace{\frac{\tilde \rho\tilde  n_x}{Pe_r}}_{(ii)}-
                  V_s\left(\pdif{}x(\tilde\rho \tilde Q_{xx})+\underbrace{\frac{1}{\eps}\pdif{}{\tilde y}( \tilde\rho \tilde Q_{xy})}_{(iii)}+\frac{1}{2}\pdif{\tilde \rho}x\right) \label{nb1_eqn}  ,     \\   \underbrace{\eps {\tilde y}\pdif{(\tilde \rho \tilde n_y)}x}_{(iv)}&=-\underbrace{\left(\frac{1}{2}-\frac{\beta}{4}\right)\tilde\rho\tilde n_x}_{(i)}-\underbrace{\frac{\tilde \rho\tilde  n_y}{Pe_r}}_{(ii)} -V_s\left(\pdif{}x(\tilde \rho \tilde Q_{xy})+\underbrace{\frac{1}{\eps}\pdif{}{\tilde y}(\tilde \rho \tilde Q_{yy})+\frac{1}{2\eps}\pdif{\tilde \rho}{\tilde y}}_{(iii)}\right)\label{nb2_eqn} .
\end{align}\label{n_bdy_layer}
\end{subequations}
  Examining Eqs.\,(\ref{n_bdy_layer}) we can identify five potential physical effects which could determine the mean orientation at leading-order their relative sizes:
 \begin{enumerate}[label=(\roman*)]
    \item shear alignment,
    \item rotational diffusion, 
    \item swimming of bacteria down vertical gradients of density and angular distribution,
    \item horizontal transport of oriented swimmers from upstream.
\end{enumerate} 
Explicitly, these effects (i-iv)  appear Eqs.\,(\ref{n_bdy_layer}) at locations we highlight with underbraces.
All other terms in Eqs.\,(\ref{n_bdy_layer}) are asymptotically sub-dominant so could not contribute at leading-order.  A balance between effects (i) and (ii) alone leads to the homogenous solution found in the outer region in Section \ref{outer-section}. Therefore other terms from Eqs.\,(\ref{n_bdy_layer}) must contribute to explain the emergent effects observed in the agent-based simulations of bacteria moving towards and adhering to the wall. For $V_s\ll1$, we consider the distinguished limit in which a balance between mechanisms (i), (ii), and (iii) yields\begin{align}
   \tilde \rho  |\bm{\tilde n}|\sim \frac{\tilde \rho|\bm{\tilde n}|}{\Pe_r}\sim \frac{V_s}{\eps}\pdif{\tilde \rho}{\tilde y}\label{LO-balance} \text{ as } \eps \to 0.
\end{align}
The scaling \eqref{LO-balance} means that effect (iv), the horizontal advection in Eqs.\,(\ref{n_bdy_layer}), is subdominant and does not contribute at leading-order. 
Using the scaling \eqref{LO-balance} in \eqref{rhob_eqn} we can now determine how the bacterial density evolves. Since both components of the mean orientation vector are of the same asymptotic order, the dominant balance in \eqref{rhob_eqn} is between horizontal advection and vertical swimming, which generates the leading-order equation:
\begin{align}
   \eps {\tilde y}\pdif{\tilde\rho}x+\frac{V_s}{\eps}\pdif{(\tilde\rho \tilde n_y)}{\tilde y}=0.\label{rhob_eqn-scalng}
\end{align}
This balance between vertical swimming and the weak shear flow tells us that we seek a solution in which there is only a small bias in the bacterial orientation in the boundary layer: $|\bm{\tilde{n}}|\to 0 $ as $\eps \to 0$.

Finally, we determine the nematic order tensor $\tilde{\bm{Q}}$ components  $\tilde Q_{xx}$, $\tilde Q_{xy}$ and $\tilde Q_{yy}$. We consider the following rescaled equations derived from \eqref{q_eqnS} using Eqs.\,(\ref{nabla}), (\ref{Q-bdy-T}), (\ref{Q-bdy-I}) and (\ref{q_components}):
\begin{subequations}
    \begin{align}
\eps {\tilde y}\pdif{(\tilde\rho \tilde Q_{xx})}x&=-\frac{V_s}{4}\left( 5\dfrac{\partial (\tilde \rho \tilde n_x)}{\partial x}+\underbrace{ \dfrac{3}{\eps}\dfrac{\partial(\tilde \rho \tilde n_y)}{\partial \tilde y}}_{(c)}\right)+\underbrace{\tilde \rho \tilde{Q}_{xy}}_{(a)}-\underbrace{\frac{4\tilde \rho\tilde  Q_{xx}}{Pe_r} }_{(b)}  \label{qb1_eqn},  \\
\eps {\tilde y}\pdif{(\tilde\rho \tilde Q_{xy})}x&= -\frac{V_s}{4}\left(\dfrac{\partial (\tilde \rho \tilde n_y)}{\partial x}+ \underbrace{\dfrac{ 1}{\eps}\dfrac{\partial(\tilde \rho \tilde n_x)}{\partial \tilde y}}_{(c)}\right)+\underbrace{\frac{\beta\tilde\rho}{4}+\frac{1}{2} \tilde\rho \tilde Q_{xx} (\beta-1)+\frac{1}{2} \tilde\rho \tilde Q_{yy} (\beta+1)}_{(a)}-\underbrace{\frac{4\tilde \rho\tilde  Q_{xy}}{Pe_r} }_{(b)} \label{qb2_eqn},  \\
              \eps {\tilde y}\pdif{(\tilde\rho \tilde Q_{yy})}x&=-\frac{V_s}{4}\left( 3\dfrac{\partial (\tilde \rho \tilde n_x)}{\partial x}+ \underbrace{\dfrac{5}{\eps}\dfrac{\partial(\tilde \rho \tilde n_y)}{\partial \tilde y}}_{(c)}\right)-\underbrace{\tilde\rho \tilde Q_{xy}}_{(a)}-\underbrace{\frac{4 \tilde \rho\tilde  Q_{yy}}{Pe_r}   }_{(b)}. \label{qb3_eqn} 
\end{align}
\end{subequations}
The underbraces denote the three potential mechanisms which could determine the leading-order equation nematic order tensor:
\begin{enumerate}[label=(\alph*)
]\item shear alignment,
    \item rotational diffusion, 
    \item vertical gradients in the mean orientation vector.
    \end{enumerate}
    The scaling of the mean-orientation vector, \eqref{LO-balance}, combined with the balance in the density transport equation \eqref{rhob_eqn-scalng}, means that effect (c) contributes at $O(V_s)$ and therefore subdominant to effects (a) and (b).
    We can conclude that the nematic order tensor is determined by shear alignment (a) and rotational diffusion (b) as in the outer region, see \S \ref{outer-section}. Hence the leading-order solution in the boundary layer is therefore equal to \eqref{outer_QS}: 
    \begin{align}
   \bm{\tilde{Q}}=\frac{\beta \Pe_r^2}{4(16+\Pe_r^2)}\begin{pmatrix}
       Pe_r & 4
       \\ 4 & -Pe_r
   \end{pmatrix}.   \label{inner_Q}
\end{align}
We can then state the leading-order equations for the mean-orientation vector components 
\begin{align}  
  0&=\left(\frac{1}{2}+\frac{\beta}{4}\right)\tilde \rho \tilde n_y-\frac{\tilde \rho\tilde  n_x}{Pe_r}-\frac{V_s}{\eps}\pdif{}{\tilde y}(\tilde \rho \tilde Q_{xy}),\quad      0=-\left(\frac{1}{2}-\frac{\beta}{4}\right)\tilde\rho\tilde n_x-\frac{\tilde \rho\tilde  n_y}{Pe_r} -\frac{V_s}{\eps}\pdif{ }{\tilde y}\left(\frac{1}{2}\tilde \rho+\tilde \rho \tilde Q_{yy}\right),\label{nb2_eqn0} 
\end{align}
which can be solved algebraically to determine the mean orientation vector components:
\begin{align}
     \tilde{\rho} \tilde{n}_{x}\sim-\frac{2\Pe_rV_s(8\tilde Q_{xy}+(2+\beta)(1+2\tilde Q_{yy})\Pe_r)}{\eps(16+(4-\beta^2)\Pe_r^2)}\pdif{\tilde{\rho}}{\tilde{y}}, \quad \tilde{\rho} \tilde{n}_{y}\sim-\frac{ 4\Pe_rV_s(2+(\beta-2)\Pe_r\tilde Q_{xy}+4\tilde Q_{yy})}{\eps(16+(4-\beta^2)\Pe_r^2)}\pdif{\tilde{\rho}}{\tilde{y}}.
\end{align}
Substituting the nematic order tensor components $\tilde Q_{xy},\tilde Q_{yy}$ we have, (11) :\begin{align}
     \tilde{\rho} \tilde{n}_{x}\sim-\frac{\Pe_r^2V_s(64+48\beta+(4-\beta^2)\Pe_r^2)}{\eps(16+\Pe_r^2)(16+(4-\beta^2)\Pe_r^2)}\pdif{\tilde{\rho}}{\tilde{y}}, \quad \tilde{\rho} \tilde{n}_{y}\sim-\frac{ 4\Pe_rV_s(32+(2-\beta)(1-\beta)\Pe_r^2)}{\eps(16+\Pe_r^2)(16+(4-\beta^2)\Pe_r^2)}\pdif{\tilde{\rho}}{\tilde{y}},\label{polar-bdyS}
\end{align}
Finally we can insert solution \eqref{polar-bdyS} into the equation for bacterial density \eqref{rhob_eqn-scalng} giving an equation in density alone, (12) :
  \begin{align}
\tilde{y}\pdif{\tilde{\rho}}x-\frac{1}{\eps^3\Pe_{\text{eff}}}\spdif{\tilde{\rho}}{\tilde{y}}&=0, \qquad \text{where } \Pe_{\text{eff}}=\frac{ (16+\Pe_r^2)(16+(4-\beta^2)\Pe_r^2)}{4\Pe_rV_s^2(32+(2-\beta)(1-\beta)\Pe_r^2)}, \label{LévêqueS}
\end{align}
which defines both the effective P\'eclet number $\Pe_{\text{eff}} $ and the boundary layer thickness $\eps \tilde{y} \sim (x\Pe_{\text{eff}})^{1/3}$.
For spherical bacteria ($\beta=0)$ the effective P\'eclet number simplifies to 
  \begin{align}
\Pe_{\text{eff}}=\frac{(4+\Pe_r^2)}{2\Pe_rV_s^2}. 
\end{align}

\subsection{Numerical solution of the Fokker-Planck equation for angular distribution}\label{S4}
In the agent-based simulations we initialise the bacteria with orientations sampled from the diffusive Jeffery-orbit distribution $\Psi(\theta)$, where $\theta$ defines the angle the orientation vector $\bm{s}$ makes with the $x$-axis. This distribution is determined by seeking steady, spatially uniform solutions to the Fokker-Plank equation, \eqref{FP}, namely $\Psi(\theta)$ satisfies 
\begin{align}
            \pdif{}\theta\left(v_\theta\Psi\right)-\frac{1}{Pe_r}\spdif{\Psi}\theta
       =0,\label{FP_t} \quad v_\theta=\frac{1}{2}(1-\beta\cos(2\theta)),
\end{align}
where $v_\theta$ is the bacterium's angular velocity in terms of $\theta$. \eqref{FP_t} is solved imposing periodicity of $\Psi$ in the domain $\theta\in[0,2\pi)$ and that the integral of the $\Psi$ in the interval $\theta\in[0,2\pi)$ is one. We use a truncated Fourier series solution for \eqref{FP_t} derived in \cite{talbot2024exploring}:
\begin{align}
    \Psi=\sum_{n=0}^N(a_n\cos(2n\theta)+b_n\sin(2n\theta)).\label{series}
\end{align}
The coefficients $(a_n,b_n)$ are defined by a series of $2N$ coupled linear equations:
\begin{subequations}
    \begin{align}
    2b_n-\beta(b_{n-1}+b_{n+1})&=\frac{8n}{\Pe_r}a_n,\\
        -2a_n+\beta((1+\delta_{k,1})a_{n-1}+a_{n+1})&=\frac{8n}{\Pe_r}b_n.
\end{align}\label{coefs}
\end{subequations}
This system is solved applying $a_{N+1},b_{N+1}=0$, $b_0=0$, $a_0=1/(2\pi)$, the latter condition ensures that the integral of $\Psi$ over $\theta\in[0,2\pi)$ is equal to one. Periodicity of the PDF is ensured by the Fourier series solution.  In our agent-based simulations we truncate series \eqref{series} at $N=30.$

\subsection{Accuracy of continuum model closure}\label{validity}
\begin{figure}
    \centering
    \includegraphics[width=0.9\linewidth]{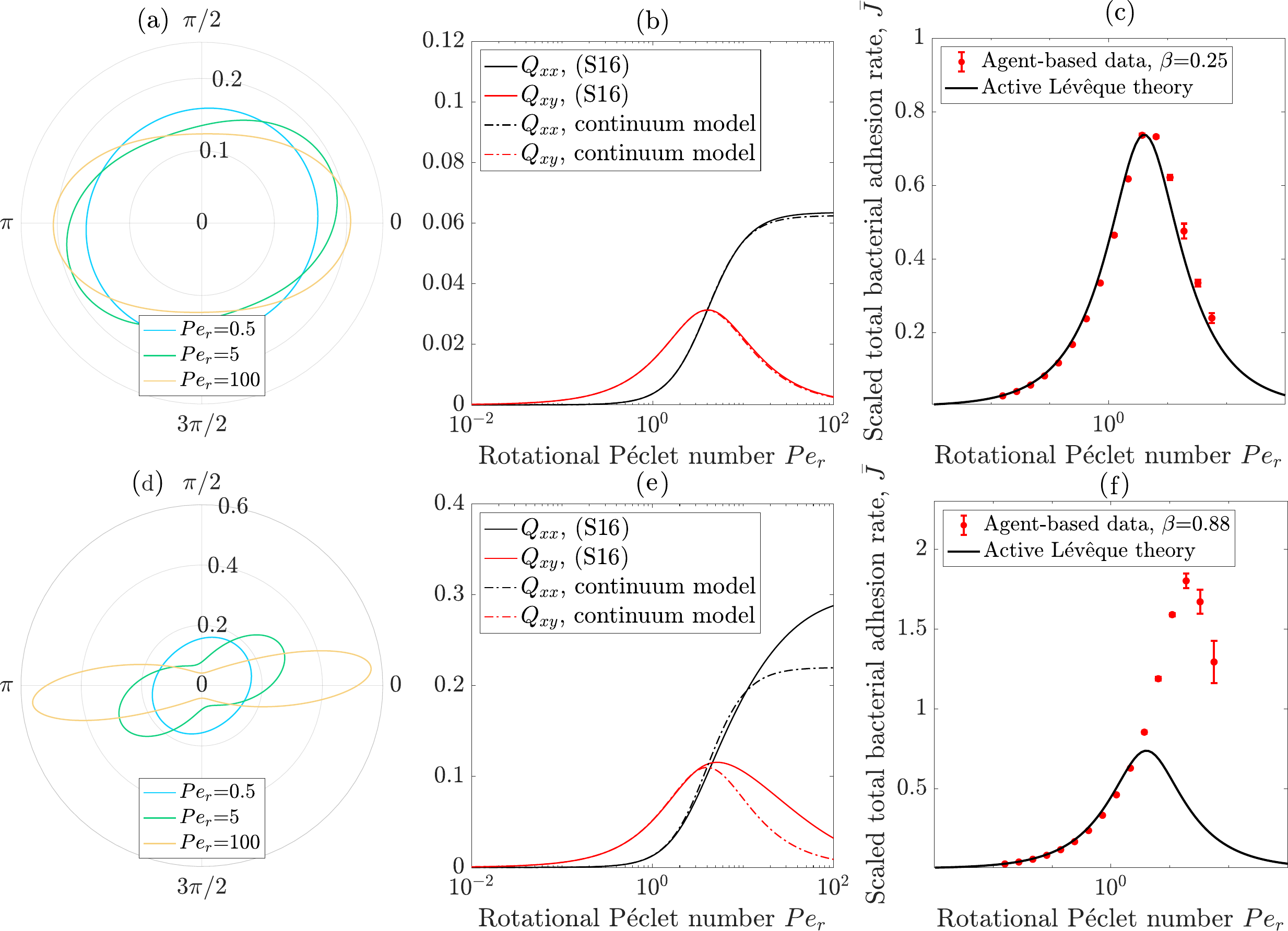}
    \caption{Angular distribution of the bacteria and the validity of the continuum model for two different bacteria shapes: $\beta=0.25$ (a--c), $\beta=0.88$ (d--f).  (a) Numerical solution \eqref{series} for $\beta=0.25$ at varying $\Pe_r$. (b) Comparison of numerical solution \eqref{series} to model solution \eqref{outer_QS} with $\beta=0.25$ shows good agreement for all $\Pe_r$.  (c) Agent-based flux is well approximated by scaled model flux $\bar{J}$ for $\beta=0.25$.  (d) Numerical solution of \eqref{series} for $\beta=0.88$ at varying $\Pe_r$.  (e) Comparison of numerical solution \eqref{series} to model solution \eqref{outer_QS} for varying $\Pe_r$ ($\beta=0.88$), shows inaccuracy of the model at large $\Pe_r$. (f) Agent-based flux is no longer well approximated by scaled model flux $\bar{J}$ for $\beta=0.88$ at large $\Pe_r$.  In all cases we truncate series \eqref{series} at $N=30.$}
    \label{fig:Jeff}
\end{figure}

\begin{figure}
    \centering
    \includegraphics[width=0.7\linewidth]{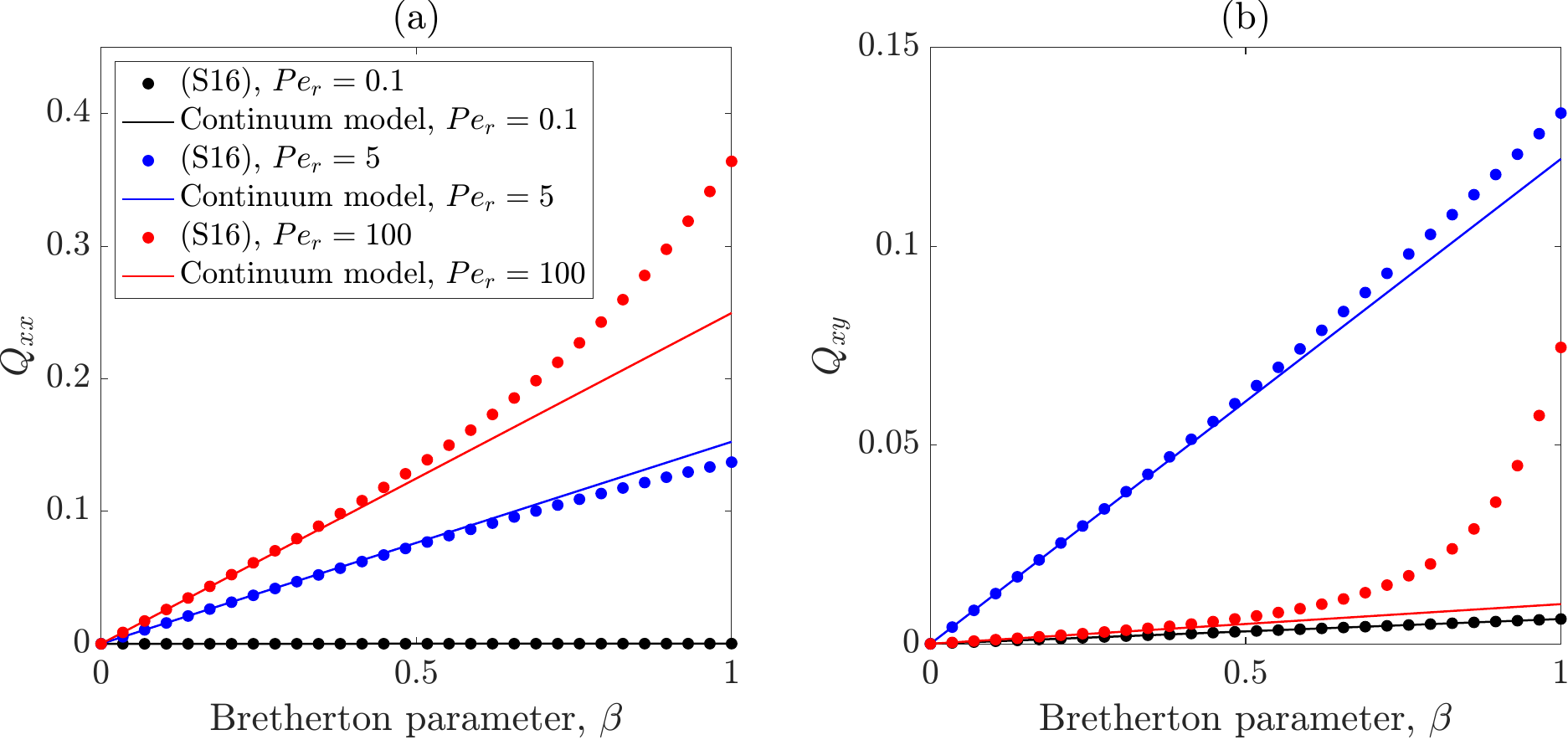}
    \caption{Comparison of numerical solution \eqref{series} to model solution \eqref{outer_Q} for $\beta \in[0,1)$. In all cases we truncate series \eqref{series} at $N=30.$ (a) $Q_{xx}$ and (b) $Q_{xy}$. }
    \label{fig:beta}
\end{figure}

In this section we quantify the accuracy of the continuum model closure \eqref{harmonic_closure}.  We first  compare the $\bm{Q}$-tensor solution \eqref{outer_QS} to direct calculation of $\bm{Q}$ using definition \eqref{definition} combined with $\psi=\Psi(\theta)$ from Eqs.\,(\ref{series})-(\ref{coefs}).
Numerical solutions of Eqs.\,(\ref{series})-(\ref{coefs}) with $N=30$ demonstrate that, as expected, rotational diffusion smooths the angular distribution of bacteria with $\Psi(\theta)$ approaching a uniform distribution for $\Pe_r\to0$ and approaching the classical Jeffery-orbit distribution for $\Pe_r\to\infty$, as shown in Fig.\,\ref{fig:Jeff}a ($\beta=0.25$) and Fig.\,\ref{fig:Jeff}d ($\beta=0.88$).
Comparison of the components of the $\bm{Q}$-tensor solution \eqref{outer_QS}
with direct calculation of $\bm{Q}$ from the series solution \eqref{series} using definition \eqref{definition} are shown in Fig.\,\ref{fig:Jeff}b ($\beta=0.25$) and Fig.\,\ref{fig:Jeff}e ($\beta=0.88$).
The solution \eqref{outer_QS} is a highly accurate approximation of $\Psi$ for modestly elongated bacteria for all  $\Pe_r\in[10^{-2},100]$ and for highly elongated bacteria at small $\Pe_r$. However, \eqref{outer_QS} fails to capture the distribution quantitively for highly elongated bacteria with large $\Pe_r$, because $\Psi$ cannot be accurately approximated by the truncated continuum model closure \eqref{harmonic_closure}, see Fig.\,\ref{fig:Jeff}e. The accuracy of the harmonic closure, \eqref{harmonic_closure} for modestly elongated bacteria allows the our active Lévêque theory to accurately predict bacterial adhesion according to  Eq.\,(16), see Fig.\,\ref{fig:Jeff}c. However, the inability of the closure to capture the angular distribution of highly elongated bacteria at large $\Pe_r$ is reflected in the disagreement between the agent based flux and the predicted flux using Eq.\,(16) for $\beta=0.88$ in Fig.\,\ref{fig:Jeff}f. A comparison of the $\bm{Q}$-tensor solution \eqref{outer_QS}
with direct calculation of $\bm{Q}$ from the series solution \eqref{series} for the full range of Bretherton parameters $\beta\in[0,1)$ is shown in Fig.\,\ref{fig:beta}. In the main text we use $\beta \in[0,0.4]$ for which \eqref{outer_QS} is accurate for all rotational P\'eclet numbers $\Pe_r\in[10^{-2},100]$.

\end{document}